\documentclass[%
superscriptaddress,
    preprint,
prl,
]{revtex4-2}

\usepackage{graphicx}% Include figure files
\usepackage{dcolumn}% Align table columns on decimal point
\usepackage{bm}% bold math
\usepackage{siunitx}

\begin{document}

\title{Mechanism of electrical switching of ultra-thin CoO/Pt bilayers}

\author{C. Schmitt}
 \email{Christin.Schmitt@Uni-Mainz.de}
 \affiliation{Institute of Physics, Johannes Gutenberg University Mainz, 55099 Mainz, Germany}
 \author{A. Rajan}
 \affiliation{Institute of Physics, Johannes Gutenberg University Mainz, 55099 Mainz, Germany}
  \author{G. Beneke}
 \affiliation{Institute of Physics, Johannes Gutenberg University Mainz, 55099 Mainz, Germany}
  \author{A. Kumar}
 \affiliation{Institute of Physics, Johannes Gutenberg University Mainz, 55099 Mainz, Germany}
  \author{T. Sparmann}
 \affiliation{Institute of Physics, Johannes Gutenberg University Mainz, 55099 Mainz, Germany}
\author{H. Meer}
 \affiliation{Institute of Physics, Johannes Gutenberg University Mainz, 55099 Mainz, Germany}
  \author{B. Bednarz}
 \affiliation{Institute of Physics, Johannes Gutenberg University Mainz, 55099 Mainz, Germany}
\author{R. Ramos}
 \affiliation{WPI-Advanced Institute for Materials Research, Tohoku University, Sendai 980-8577, Japan}
 \affiliation{Present address: Centro Singular de Investigación en Química Biolóxica e Materiais Moleculares (CiQUS), Departamento de Química-Física, Universidade de Santiago de Compostela, Santiago de Compostela 15782, Spain}
 \author{M. Angel Niño}
 \affiliation{ALBA Synchrotron Light Facility, Cerdanyola del Vallés, 08290 Barcelona, Spain}
  \author{M. Foerster}
 \affiliation{ALBA Synchrotron Light Facility, Cerdanyola del Vallés, 08290 Barcelona, Spain}
 \author{E. Saitoh}
\affiliation{WPI-Advanced Institute for Materials Research, Tohoku University, Sendai 980-8577, Japan}
 \affiliation{Institute for Materials Research, Tohoku University, Sendai 980-8577, Japan}
 \affiliation{The Institute of AI and Beyond, The University of Tokyo, Tokyo 113-8656, Japan}
 \affiliation{Center for Spintronics Research Network, Tohoku University, Sendai 980-8577, Japan}
 \affiliation{Department of Applied Physics, The University of Tokyo, Tokyo 113-8656, Japan}
\author{M. Kläui}
\email{Klaeui@Uni-Mainz.de}
 \affiliation{Institute of Physics, Johannes Gutenberg University Mainz, 55099 Mainz, Germany}
 \affiliation{Graduate School of Excellence Materials Science in Mainz, 55128 Mainz, Germany}

\date{\today}% It is always \today, today,

\begin{abstract}
We study current-induced switching of the Néel vector in CoO/Pt bilayers to understand the underlaying antiferromagnetic switching mechanism. Surprisingly, we find that for ultra-thin CoO/Pt bilayers electrical pulses along the same path can lead to an increase or decrease of the spin Hall magnetoresistance signal, depending on the current density of the pulse. By comparing the results of these electrical measurements to XMLD-PEEM imaging of the antiferromagnetic domain structure before and after the application of current pulses, we reveal the reorientation of the Néel vector in ultra-thin CoO(\SI{4}{nm}). This allows us to determine that even opposite resistance changes can result from a thermomagnetoelastic switching mechanism. Importantly, our spatially resolved imaging shows that regions where the current pulses are applied and regions further away exhibit different switched spin structures, which can be explained by a spin-orbit torque based switching mechanism that can dominate in very thin films.

\end{abstract}

\maketitle

%\tableofcontents
\newpage

\newpage
\section{\label{sec:intro}Introduction\protect}
Antiferromagnets have recently attracted attention as possible materials for future spintronic devices \cite{Baltz2018}. This is because of their unique properties that may allow to overcome the limitations of current systems, which conventionally use ferromagnets as active elements. Due to their zero net magnetic moment AFMs allow for higher bit packing densities and enhanced stability against external magnetic fields \cite{meer2023antiferromagnetic, loth2012bistability}. Furthermore, AFMs potentially allow for ultrafast applications due to their internal dynamics with resonant frequencies in the THz range \cite{kampfrath2011coherent}. Insulating antiferromagnetic materials are of particular interest for the development of low power devices, as their low damping allows for the transport of pure spin currents over long distances \cite{lebrun2018tunable, das}.
However, the absence of a net magnetic moment in antiferromagnets makes efficient reading and writing of magnetic information challenging. It has been established that for insulating AFMs, electrical currents in an adjacent heavy metal layer can be used to read out the orientation of the Néel vector via the spin Hall magnetoresistance (SMR) \cite{hoogeboom2017negative, baldrati2018full}. Electrical writing of information has been reported via short current pulses which can induce a reorientation of the antiferromagnetic order, both in metallic AFMs \cite{wadley2016electrical, bodnar2018writing, lytvynenko2022current} and bilayers of insulating AFMs and heavy metals \cite{moriyama2018spin, chen2018antidamping, baldrati2019mechanism}. The switching mechanism in the latter case is, however, being debated in terms of origin and efficiency. Some studies suggest that the switching is dominated by a thermomagnetoelastic mechanism, especially in materials with large magnetostriction \cite{zhang2019quantitative, meer2020direct, baldrati2020efficient}. Other studies have proposed that the switching can also be driven by spin-orbit torques (SOTs) \cite{moriyama2018spin, chen2018antidamping, baldrati2019mechanism, gray2019spin, cheng2020electrical, zhang2022control}. Initially, it was surprising to find current-induced switching for AFM films with tens of nm thickness given the interfacial nature of the SOTs. One can expect that for thicker films the switching is dominated by a thermomagnetoelastic switching mechanism while spin-orbit torques are expected to play a key role in thin films. The dampinglike spin-orbit torque effective fields, generated by electrical current pulses in AFMs, scale with $1/d$, where $d$ is the thickness of the AFM layer \cite{baldrati2019mechanism}. Thus, for electrical switching experiments on insulating antiferromagnetic materials very thin CoO films are a good choice. CoO is a suitable candidate for switching experiments because it is a collinear compensated AFM with a Néel temperature of $T_{\mathrm{N}} = \SI{291}{K}$ in the bulk \cite{roth1958magnetic, uchida1964magnetic, saito1966x}, easily accessible with commonly used experimental setups. When CoO is grown on MgO(001) substrates under compressive strain (lattice mismatch 1.1 \%) \cite{csiszar2005controlling, zhu2014antiferromagnetic}, this can induce $T_\mathrm{N}$ around room temperature and a fourfold in-plane magnetic configuration with two orthogonal stable states \cite{cao2011temperature, baldrati2020efficient, grzybowski2023antiferromagnetic}. Such a configuration is ideal for electrical switching experiments to write magnetic information and the electrical reading of the Néel vector via SMR \cite{chen2018antidamping, nakayama2013spin, baldrati2018full}. However, to efficiently make use of the advantages of insulating AFMs in general and CoO in particular as active elements in spintronic devices, the switching mechanism must be identified. One needs to understand if transport measurements suffice to identify the mechanism and analyze if competing mechanisms are present simultaneously.\\
In this work, we investigate current-induced switching of ultra-thin CoO/Pt bilayers and identify the switching mechanism by combining electrical measurements and imaging of the magnetic domain structure for different pulse current densities. First, we show that in ultra-thin CoO films different switching signs can be observed in electrical SMR measurements. We demonstrate by direct imaging that the electrically observed switching of both signs can be explained by a single mechanism that is consistent with a thermomagnetoelastic process. However, by analyzing the switching in different areas of the sample, we can identify regions where the switching can be explained by an additional SOT-based mechanism. In this way, we demonstrate that transport measurements are not sufficient to identify the switching mechanism, but direct imaging of the antiferromagnetic domain structure in ultra-thin samples reveales that both thermomagnetoelastic switching and SOT-switching together result in the observed current induced spin structure changes.

\section{\label{sec:results}Results\protect}

\begin{figure}
 \includegraphics[width=16cm]{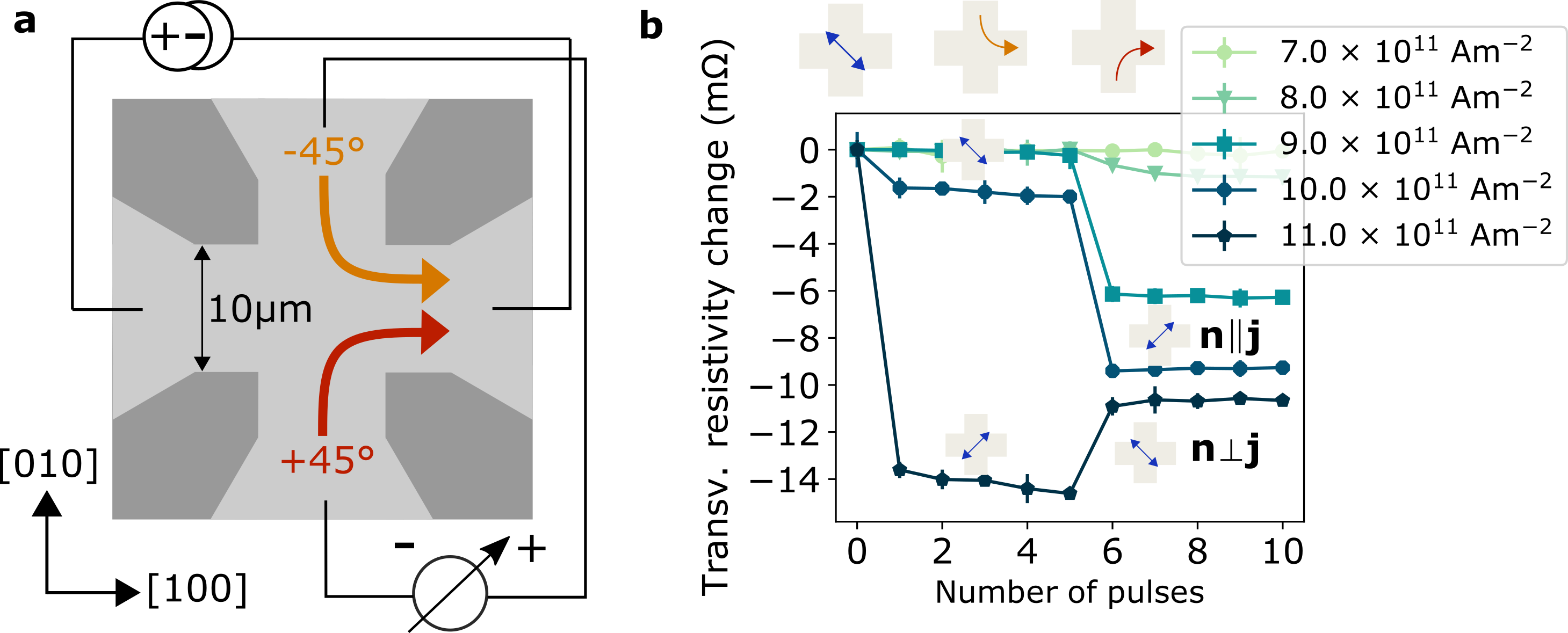}
 \caption{(a) Device and contact scheme used for the electrical measurements. (b) Current density-dependent switching of the Néel vector $\mathbf{n}$ with respect to the current direction $\mathbf{j}$ in CoO($\SI{4}{nm}$)/Pt($\SI{2}{nm}$) thin-films showing two different switching regimes. The pulsing directions for -45°- (pulse 1-5) and +45°-pulses (pulse 6-10) are indicated by the orange and red arrow, accordingly. The blue arrows indicate the orientation of the Néel vector before and after the pulses. For low current densities $\mathbf{n}$ switches parallel to $\mathbf{j}$. For high current densities $\mathbf{n}$ switches perpendicular to $\mathbf{j}$.}
 \label{fig:one}
\end{figure}

To investigate the current-induced effects on the antiferromagnetic domain structure, we have grown epitaxial CoO($d$)/Pt($\SI{2}{nm}$) bilayers on MgO(001) substrates, where $d = \SI{4}{nm}$ and $\SI{2}{nm}$. Using Ar ion beam etching we have patterned Hall cross structures with the arms of the cross being aligned parallel to the [100] and [010] sample edges (hard axes). We have etched crosses with a channel width of $\SI{10}{\micro m}$. Edge pulses around the corners of the cross result in a current direction $\mathbf{j}$ in the center of the device parallel or perpendicular to the two easy axes ([110] and $[\bar{1}10]$). The device layout is shown in Fig. \ref{fig:one} (a). The layout allows for a reproducible switching between two orthogonal Néel vector orientations. $\SI{1}{ms}$ long current pulses of varying current density are applied to the Hall cross arms. The resulting orientation of the Néel vector $\mathbf{n}$ is read electrically, via the transverse SMR signal, which is proportional to the in-plane Néel vector components $n_x \cdot n_y$ \cite{baldrati2018full}. The SMR change is maximised for two states with orthogonal orientation of $\mathbf{n}$.\\
First, we investigate the switching in a CoO($\SI{4}{nm}$)/Pt($\SI{2}{nm}$) film. Before the measurements, a magnetic field $\mu_0 \mathbf{H}_{\mathrm{before}} = \SI{12}{T}$ along the [110] direction is applied to the sample to set the Néel vector $\mathbf{n}$ into a well defined starting state along the $[\bar{1}10]$ direction (blue double arrow in Fig. \ref{fig:one} (b)) and then removed. This is followed by five pulses at -45° ($[1\bar{1}0]$) and five pulses applied at +45° ($[110]$) with respect to the measurement current direction. Fig. \ref{fig:one} (b) shows that for low current densities $\mathbf{j}_{\mathrm{pulse}} \leq \SI{8e11}{Am}^{-2}$ no current-induced switching can be detected in the SMR signal. Increasing the current density to $\mathbf{j}_{\mathrm{pulse}} = \SI{9e11}{Am}^{-2}$ still does not show any switching of the Néel vector after the application of five pulses at -45°. However, the application of pulses along +45° induces a decrease of the transverse SMR signal, indicating that at least part of the Hall cross was switched with a final state of the Néel vector $\mathbf{n} \parallel \mathbf{j}$. Further increasing the current-density to $\mathbf{j}_{\mathrm{pulse}} = \SI{10e11}{Am}^{-2}$ shows an additional small decrease in the transverse resistivity for -45°-pulses, which is consistent with a reorientation of $\mathbf{n} \perp \mathbf{j}$. For +45°-pulses the SMR signal decreases further, indicating an opposite switching sign $\mathbf{n} \parallel \mathbf{j}$. This difference in the sign of the observed switching signal could be taken as an indication that different switching mechanisms are acting. The different switching signs become even more pronounced when increasing the current density further to $\mathbf{j}_{\mathrm{pulse}} = \SI{11e11}{Am}^{-2}$. For this current density the effect of the -45°-pulse is more pronounced and a large decrease in transverse resistivity signal can be observed after the first applied pulse. This change in the electrical signal indicates a reorientation of $\mathbf{n} \perp \mathbf{j}$. However, +45°-pulses at this current density lead to an increase of the transverse SMR signal, also indicating $\mathbf{n} \perp \mathbf{j}$ (as the Néel vector orientation was switched already by the first -45°-pulse and was oriented along [110]). Thus, the switching sign for $\mathbf{j}_{\mathrm{pulse}} = \SI{11e11}{Am}^{-2}$ is opposite to what was observed for the lower current densities $\mathbf{j}_{\mathrm{pulse}} \leq \SI{10e11}{Am}^{-2}$. From the electrical measurements we can, therefore, conclude that in ultra-thin CoO($\SI{4}{nm}$)/Pt($\SI{2}{nm}$) bilayers two electrical switching regimes with opposite final states of the Neel vector are present. This observation suggests that the different switching regimes originate from different mechanisms, dominating at different pulse current densities.\\
Since the strength of the SOTs scales inversely with the layer thickness, stronger SOT-induced switching is expected for thinner films \cite{baldrati2019mechanism}. Therefore, we also investigated an ultra-thin
CoO($\SI{2}{nm}$)/Pt($\SI{2}{nm}$) sample (details are shown in the supporting information). Just as for the $\SI{4}{nm}$ sample, two different switching regimes can be identified in transport measurements. At low current densities, one observes an orientation of the Néel vector final state parallel to the current direction ($\mathbf{n} \parallel \mathbf{j}$) and at higher current densities a perpendicular orientation ($\mathbf{n} \perp \mathbf{j}$). Beyond that, however, no further insights into the switching mechanisms in antiferromagnetic thin films can be obtained from the transport study of this sample.\\
Thus, for these ultra-thin CoO samples the open question is whether two different mechanisms cause a switching of the Néel vector. Since transport signals can suffer from issues with a spatially varying sensitivity of the Hall cross device \cite{schreiber2021magnetic} or due to non-magnetic contributions to the SMR signal, from e.g. electromigration effects \cite{schreiber2020concurrent}, imaging of the AFM domain structure is needed to correlate the effect of the electrical current pulses on the spatially resolved domain structure and to be able to draw conclusions about the acting mechanisms and origin of the different switching regimes.

\begin{figure*}
 \includegraphics[width=15cm]{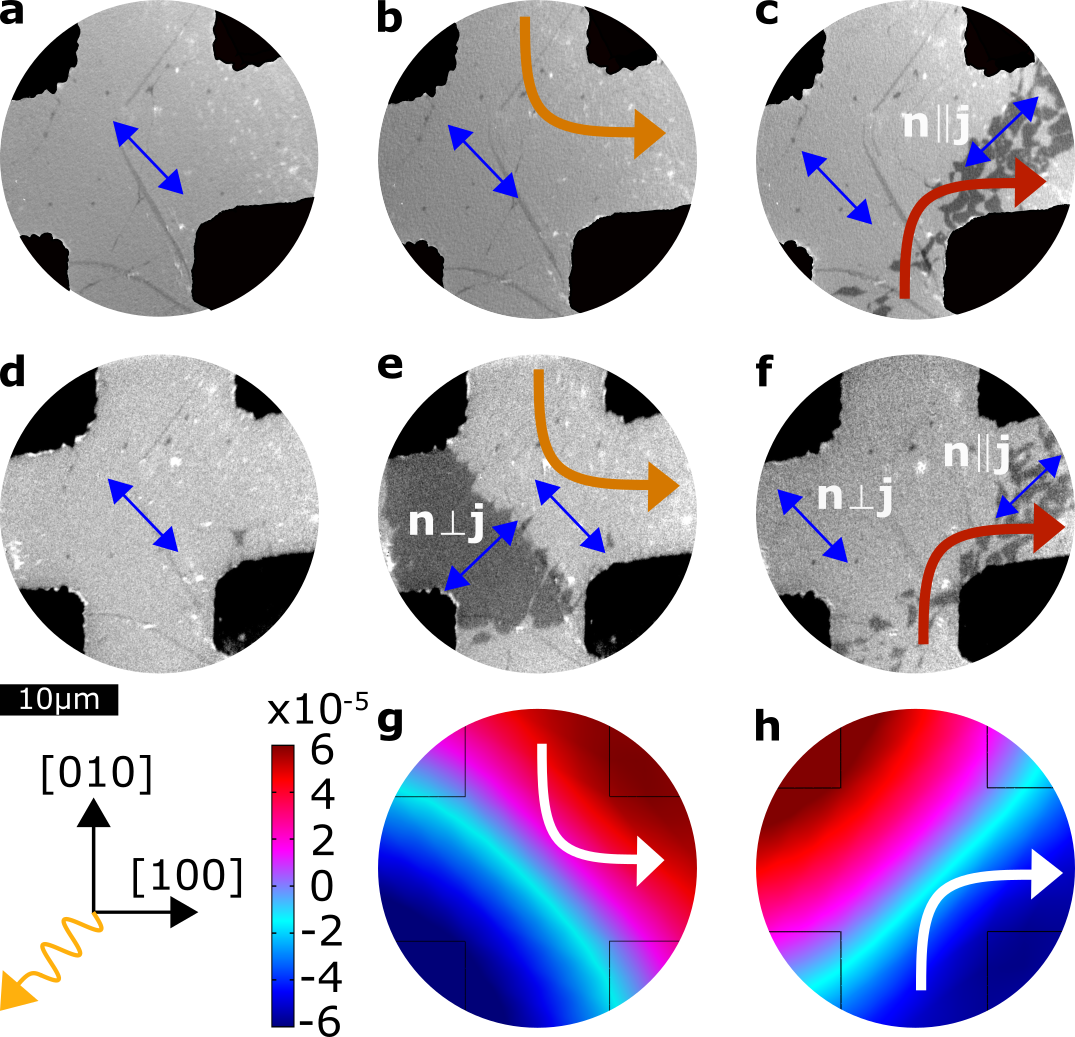}
 \caption{XMLD images of a CoO($\SI{4}{nm}$)/Pt($\SI{2}{nm}$) film. (a) The antiferromagnetic domain structure is in a saturated nearly monodomain state after the application of a magnetic field of $\SI{12}{T}$ at $\SI{300}{K}$ in the [$\bar{1}10$] direction. The Néel vector orientation is indicated by the blue double arrow. (b) Domain structure after the application of a -45°-pulse (orange) with $\mathbf{j}_{\mathrm{pulse}} = \SI{8.5e11}{Am}^{-2}$. (c) The application of a +45°-pulse induces some reorientation of the Néel vector with $\mathbf{n} \parallel \mathbf{j}$. (d) Heating the sample up above Néel temperature restores the almost monodomain state with $\mathbf{n} \parallel [110]$. (e) The application of a -45°-pulse ($\mathbf{j}_{\mathrm{pulse}} = \SI{11.5e11}{Am}^{-2}$) induces a change of the antiferromagnetic domain structure. (f) Magnetic state after a +45°-pulse with $\mathbf{j}_{\mathrm{pulse}} = \SI{11.5e11}{Am}^{-2}$. (g) Strain-profile for a pulse along the -45° direction. (h) Strain-profile for a pulse along the +45° direction.}
 \label{fig:two}
\end{figure*}

When comparing the electrical signal with the domain structure changes, we can directly conclude on which of the switching signs of the regimes is consistent with which of the proposed switching mechanisms for insulating antiferromagnetic/heavy metal bilayers \cite{meer2020direct}.\\
To image the AFM domain structure, we use photoemission electron microscopy (PEEM) exploiting the X-ray magnetic linear dichroism (XMLD) effect. We image the AFM domain structure using energy dependent XMLD-PEEM at the Co $\mathrm{L_3}$ edge with the two energies $E_1 = \SI{777.9}{eV}$ and $E_2 = \SI{779.0}{eV}$. Fig. \ref{fig:two} (a) shows the investigated Hall bar region. Before imaging the domain structure a magnetic field above spin-flop of $\mu_0 \mathbf{H} = \SI{12}{T}$ was applied to the sample at a temperature of $\SI{300}{K}$ along the [$\bar{1}10$] direction, in order to have the sample in the same starting state as for the transport measurements, with the Néel vector being oriented along the [110] direction (indicated by the blue double arrow). It needs to be noted that the dark stripes that can be observed in the center of the Hall cross structure are non-magnetic in origin, caused by inhomogeneities on the surface. However, since the values of the resistance in the Hall cross are comparable for all measured devices, we can conclude that the surface inhomogeneities do not influence the path of the current pulses. When we apply a -45°-pulse, indicated by the orange arrow in Fig. \ref{fig:two} (b), with $\mathbf{j}_{\mathrm{pulse}} = \SI{8.5e11}{Am}^{-2}$, no significant change in the magnetic domain structure is observed. This is consistent with the low current regime in the electrical measurements, where a -45°-pulse does not lead to an observable change in transverse Hall resistivity. A +45°-pulse with the same current density, on the other hand, induces a reorientation of the Néel vector in the region of the Hall cross where the current is flowing, as it is shown in Fig. \ref{fig:two} (c). This magnetic change is consistent with a switching of $\mathbf{n}\parallel \mathbf{j}$, in line with the low current density regime in the electrical measurements. Before the next current pulses were applied, the magnetic state of the sample was brought back into the almost monodomain state with $\mathbf{n} \parallel [110]$ by heating it in-situ above its Néel temperature. The subsequent application of a single pulse with $\mathbf{j}_{\mathrm{pulse}} = \SI{11.5e11}{Am}^{-2}$ along the -45° direction induces a change of the AFM domain structure in the arms which are opposite to those to which the current pulse is applied. Fig. \ref{fig:two} (e) shows the resulting magnetic state. The Néel vector in the switched region is oriented perpendicular to the current direction in the center of the device ($\mathbf{n} \perp \mathbf{j}$). This is in line with the sign of the switching observed in the high current density regime in the electrical measurements. A pulse with equal current density but perpendicular direction (+45°), as shown in Fig. \ref{fig:two} (f), erases parts of the previously nucleated domain structure and generates a magnetic state similar to that created when applying a smaller current density. Having a closer look at the created magnetic domain structure reveals that in this case not a single sign of the switching can be determined. In the arms of the cross  opposite to where the current pulse was applied the final state of the Néel vector is $\mathbf{n} \perp \mathbf{j}$ and is consistent to what the electrical measurements show. On the other hand, in the arms where the current pulse was applied the final state is $\mathbf{n} \parallel \mathbf{j}$. Thus, the microscopic measurements show that different switching behaviour can be observed in different parts of the sample and that there are also differences depending on the current density. To infer the switching mechanism, we compare the XMLD-PEEM images with the switching sign predicted for SOTs in CoO and with the difference in strain resulting from the current-induced heat for the two pulsing directions which we simulated with COMSOL. For the simulations in Fig. \ref{fig:two} (g), (h) we plot the difference between the strain along the two easy axes $\epsilon[\bar{1}10] - \epsilon[110]$. A positive strain difference, shown in red, corresponds to a larger expansion of the CoO along $[\bar{1}10]$ in contrast to $[110]$, while a negative strain difference (blue) indicates larger strain along the orthogonal direction. Fig. \ref{fig:two} (g) and (h) show the resulting strain-profiles for the pulses along the -45° and +45° direction, respectively. Comparing these strain-profiles with the XMLD-PEEM images in the low current-density regime indicates that after a -45°-pulse (Fig. \ref{fig:two} (b)), no switching is observed because the Néel vecor throughout the device is already oriented along the $[110]$ direction, the axis along which a stronger expansion of the CoO is observed in the region where the current pulse is applied (red). In the arms opposite to where the current pulse is applied the strain is in the orthogonal direction and could lead to a reorientation of the Néel vector which, however, is not observed. One reason for this could be that the heat in this area is not sufficient to trigger a switching of the Néel vector \cite{meer2020direct}. For a +45°-pulse, on the other hand, the region with stronger expansion of the CoO along $[\bar{1}10]$ (blue) is in the arms of the cross where the current pulse is applied and the heating is stronger. Here, a switching of the Néel vector can be observed.

It can be concluded, that the low current-density regime is compatible with a thermomagnetoelastic switching mechanism. Checking the high current-density regime reveals that the changes in the Néel vector are also consitent with a thermomagnetoelastic switching mechanism. For the -45°-pulse there is a reorientation of the Néel vector in the arms opposite to where the current pulse is applied, as there is more heating now (Fig. \ref{fig:two} (e)). For the +45°-pulse the resulting domain structure in the arms where the current pulse is applied switches into the $[\bar{1}10]$ direction (according to the strain in $[\bar{1}10]$ direction) and in the opposite part switches back into the $[110]$ direction, according to the strain in $[110]$ direction (red).\\
However, not all changes observed can be atributed to thermomagnetoelastic switching. The domain structure for +45°-pulses, when the change of the Néel vector occurs where the current pulses are applied, is different from that at -45°-pulses, where the change occurs in the opposite arms. For +45°-pulses, the resulting domain structure is less homogeneous and requires a different explanation. This thus demonstrates the presence of different switching mechanisms. To check if in the arms where the pulses are applied a spin-orbit torque based mechanism is present, we compare the theoretical predictions for the switching sign in the SOT based switching mechanism in CoO, which predicts a parallel orientation of the Neel vector $\mathbf{n}$ and applied current pulse $\mathbf{j}$ \cite{baldrati2020efficient}, to the experimentally observed switching sign. For a -45°-pulse we do not observe any switching where the current pulse is applied, which is in agreement with a SOT switching mechanism, since the Néel vector orientation is already parallel to the current direction in the center of the Hall cross. For the +45°-pulse we observe a switching of $\mathbf{n} \parallel \mathbf{j}$ in the region of the applied current pulse. We thus conclude that the switching is governed by a combination of a thermomagnetoelastic switching mechanism and a SOT based mechanism present in ultra-thin AFM films. 

\section{\label{sec:conclusion}Conclusion}
In conclusion, we have investigated the current-induced switching of the Néel vector in ultra-thin CoO/Pt bilayers. We find that electrical pulses along the same path can show in transport measurements opposite switching signs of the SMR signal for different current densities. However, by comparing these results with the XMLD-PEEM imaging of the domain structure, we show that the current-induced switching leading to the different sign of the SMR in transport measurements can be explained by the action of a single heat-assisted thermomagnetoelastic switching mechanism at all current densities. With this we show that the electrical data itself can be misleading and only imaging of the Néel vector state can reveal the full changes in the antiferromagnetic domain structure. In addition to the thermomagnetoelastic mechanism, we find that a SOT-based mechanism explains the difference in domain structure in the arms were the current pulse is applied and in those opposite to it. It is expected that such a SOT-based mechanism becomes relevant only for our thinner films, compared to previous reports of tens of nm thick films where only thermomagnetoelastic switching mechanisms were identified. Our results will motivate additional research into thinner films to enable using AFMs in applications where electrical reading and writing by SOTs are key.

%\begin{acknowledgments}
\section{acknowledgements}
 Experiments were performed at the CIRCE beamline at ALBA Synchrotron with the collaboration of ALBA staff. M.K. acknowledges support from the Graduate School of Excellence Materials Science in Mainz (MAINZ) DFG 266, the DAAD (Spintronics network, Project No. 57334897). We acknowledge that this work is funded by the Deutsche Forschungsgemeinschaft (DFG, German Research Foundation)—TRR 173-268565370 (Projects A01 and B02). This project has received funding from the European Union’s Horizon 2020 research and innovation programme under Grant Agreement No. 863155 (s-Nebula). A.R. and B.B. acknowledge funding from the European Union's Horizon 2020 research and innovation programme under the Marie Skłodowska-Curie Grant Agreement No. 860060 (ITN MagnEFi). This work is also supported by ERATO “Spin Quantum Rectification Project” (Grant No. JPMJER1402) and the Grant-in-Aid for Scientific Research on Innovative Area, “Nano Spin Conversion Science” (Grant No. JP26103005), Grant-in-Aid for Scientific Research (S) (Grant No. JP19H05600), Grant-in-Aid for Scientific Research (C) (Grant No. JP20K05297) from JSPS KAKENHI. R.R. also acknowledges support from the Grant RYC 2019-026915-I and the Project TED2021-130930B-I00 funded by the MCIN/AEI/10.13039/501100011033 and by the ESF investing in your future and the European Union NextGenerationEU/PRTR,  the Xunta de Galicia (ED431F 2022/04, ED431B 2021/013, Centro Singular de Investigación de Galicia Accreditation 2019-2022, ED431G 2019/03) and the European Union (European Regional Development Fund - ERDF).
%\end{acknowledgments}

\bibliography{main}% Produces the bibliography via BibTeX.

\begin{thebibliography}{30}%
\makeatletter
\providecommand \@ifxundefined [1]{%
 \@ifx{#1\undefined}
}%
\providecommand \@ifnum [1]{%
 \ifnum #1\expandafter \@firstoftwo
 \else \expandafter \@secondoftwo
 \fi
}%
\providecommand \@ifx [1]{%
 \ifx #1\expandafter \@firstoftwo
 \else \expandafter \@secondoftwo
 \fi
}%
\providecommand \natexlab [1]{#1}%
\providecommand \enquote  [1]{``#1''}%
\providecommand \bibnamefont  [1]{#1}%
\providecommand \bibfnamefont [1]{#1}%
\providecommand \citenamefont [1]{#1}%
\providecommand \href@noop [0]{\@secondoftwo}%
\providecommand \href [0]{\begingroup \@sanitize@url \@href}%
\providecommand \@href[1]{\@@startlink{#1}\@@href}%
\providecommand \@@href[1]{\endgroup#1\@@endlink}%
\providecommand \@sanitize@url [0]{\catcode `\\12\catcode `\$12\catcode
  `\&12\catcode `\#12\catcode `\^12\catcode `\_12\catcode `\%12\relax}%
\providecommand \@@startlink[1]{}%
\providecommand \@@endlink[0]{}%
\providecommand \url  [0]{\begingroup\@sanitize@url \@url }%
\providecommand \@url [1]{\endgroup\@href {#1}{\urlprefix }}%
\providecommand \urlprefix  [0]{URL }%
\providecommand \Eprint [0]{\href }%
\providecommand \doibase [0]{https://doi.org/}%
\providecommand \selectlanguage [0]{\@gobble}%
\providecommand \bibinfo  [0]{\@secondoftwo}%
\providecommand \bibfield  [0]{\@secondoftwo}%
\providecommand \translation [1]{[#1]}%
\providecommand \BibitemOpen [0]{}%
\providecommand \bibitemStop [0]{}%
\providecommand \bibitemNoStop [0]{.\EOS\space}%
\providecommand \EOS [0]{\spacefactor3000\relax}%
\providecommand \BibitemShut  [1]{\csname bibitem#1\endcsname}%
\let\auto@bib@innerbib\@empty
%</preamble>
\bibitem [{\citenamefont {Baltz}\ \emph {et~al.}(2018)\citenamefont {Baltz},
  \citenamefont {Manchon}, \citenamefont {Tsoi}, \citenamefont {Moriyama},
  \citenamefont {Ono},\ and\ \citenamefont {Tserkovnyak}}]{Baltz2018}%
  \BibitemOpen
  \bibfield  {author} {\bibinfo {author} {\bibfnamefont {V.}~\bibnamefont
  {Baltz}}, \bibinfo {author} {\bibfnamefont {A.}~\bibnamefont {Manchon}},
  \bibinfo {author} {\bibfnamefont {M.}~\bibnamefont {Tsoi}}, \bibinfo {author}
  {\bibfnamefont {T.}~\bibnamefont {Moriyama}}, \bibinfo {author}
  {\bibfnamefont {T.}~\bibnamefont {Ono}},\ and\ \bibinfo {author}
  {\bibfnamefont {Y.}~\bibnamefont {Tserkovnyak}},\ }\bibfield  {title}
  {\bibinfo {title} {{Antiferromagnetic spintronics}},\ }\href
  {https://doi.org/10.1103/RevModPhys.90.015005} {\bibfield  {journal}
  {\bibinfo  {journal} {Reviews of Modern Physics}\ }\textbf {\bibinfo {volume}
  {\textbf{90}}},\ \bibinfo {pages} {015005} (\bibinfo {year}
  {2018})}\BibitemShut {NoStop}%
\bibitem [{\citenamefont {Meer}\ \emph {et~al.}(2023)\citenamefont {Meer},
  \citenamefont {Gomonay}, \citenamefont {Wittmann},\ and\ \citenamefont
  {Kläui}}]{meer2023antiferromagnetic}%
  \BibitemOpen
  \bibfield  {author} {\bibinfo {author} {\bibfnamefont {H.}~\bibnamefont
  {Meer}}, \bibinfo {author} {\bibfnamefont {O.}~\bibnamefont {Gomonay}},
  \bibinfo {author} {\bibfnamefont {A.}~\bibnamefont {Wittmann}},\ and\
  \bibinfo {author} {\bibfnamefont {M.}~\bibnamefont {Kläui}},\ }\bibfield
  {title} {\bibinfo {title} {{Antiferromagnetic insulatronics: Spintronics in
  insulating 3d metal oxides with antiferromagnetic coupling}},\ }\href@noop {}
  {\bibfield  {journal} {\bibinfo  {journal} {Applied Physics Letters}\
  }\textbf {\bibinfo {volume} {122}},\ \bibinfo {pages} {080502} (\bibinfo
  {year} {2023})}\BibitemShut {NoStop}%
\bibitem [{\citenamefont {Loth}\ \emph {et~al.}(2012)\citenamefont {Loth},
  \citenamefont {Baumann}, \citenamefont {Lutz}, \citenamefont {Eigler},\ and\
  \citenamefont {Heinrich}}]{loth2012bistability}%
  \BibitemOpen
  \bibfield  {author} {\bibinfo {author} {\bibfnamefont {S.}~\bibnamefont
  {Loth}}, \bibinfo {author} {\bibfnamefont {S.}~\bibnamefont {Baumann}},
  \bibinfo {author} {\bibfnamefont {C.~P.}\ \bibnamefont {Lutz}}, \bibinfo
  {author} {\bibfnamefont {D.}~\bibnamefont {Eigler}},\ and\ \bibinfo {author}
  {\bibfnamefont {A.~J.}\ \bibnamefont {Heinrich}},\ }\bibfield  {title}
  {\bibinfo {title} {{Bistability in atomic-scale antiferromagnets}},\
  }\href@noop {} {\bibfield  {journal} {\bibinfo  {journal} {Science}\ }\textbf
  {\bibinfo {volume} {335}},\ \bibinfo {pages} {196} (\bibinfo {year}
  {2012})}\BibitemShut {NoStop}%
\bibitem [{\citenamefont {Kampfrath}\ \emph {et~al.}(2011)\citenamefont
  {Kampfrath}, \citenamefont {Sell}, \citenamefont {Klatt}, \citenamefont
  {Pashkin}, \citenamefont {Mährlein}, \citenamefont {Dekorsy}, \citenamefont
  {Wolf}, \citenamefont {Fiebig}, \citenamefont {Leitenstorfer},\ and\
  \citenamefont {Huber}}]{kampfrath2011coherent}%
  \BibitemOpen
  \bibfield  {author} {\bibinfo {author} {\bibfnamefont {T.}~\bibnamefont
  {Kampfrath}}, \bibinfo {author} {\bibfnamefont {A.}~\bibnamefont {Sell}},
  \bibinfo {author} {\bibfnamefont {G.}~\bibnamefont {Klatt}}, \bibinfo
  {author} {\bibfnamefont {A.}~\bibnamefont {Pashkin}}, \bibinfo {author}
  {\bibfnamefont {S.}~\bibnamefont {Mährlein}}, \bibinfo {author}
  {\bibfnamefont {T.}~\bibnamefont {Dekorsy}}, \bibinfo {author} {\bibfnamefont
  {M.}~\bibnamefont {Wolf}}, \bibinfo {author} {\bibfnamefont {M.}~\bibnamefont
  {Fiebig}}, \bibinfo {author} {\bibfnamefont {A.}~\bibnamefont
  {Leitenstorfer}},\ and\ \bibinfo {author} {\bibfnamefont {R.}~\bibnamefont
  {Huber}},\ }\bibfield  {title} {\bibinfo {title} {{Coherent terahertz control
  of antiferromagnetic spin waves}},\ }\href@noop {} {\bibfield  {journal}
  {\bibinfo  {journal} {Nature Photonics}\ }\textbf {\bibinfo {volume}
  {\textbf{5}}},\ \bibinfo {pages} {31} (\bibinfo {year} {2011})}\BibitemShut
  {NoStop}%
\bibitem [{\citenamefont {Lebrun}\ \emph {et~al.}(2018)\citenamefont {Lebrun},
  \citenamefont {Ross}, \citenamefont {Bender}, \citenamefont {Qaiumzadeh},
  \citenamefont {Baldrati}, \citenamefont {Cramer}, \citenamefont {Brataas},
  \citenamefont {Duine},\ and\ \citenamefont {Kläui}}]{lebrun2018tunable}%
  \BibitemOpen
  \bibfield  {author} {\bibinfo {author} {\bibfnamefont {R.}~\bibnamefont
  {Lebrun}}, \bibinfo {author} {\bibfnamefont {A.}~\bibnamefont {Ross}},
  \bibinfo {author} {\bibfnamefont {S.}~\bibnamefont {Bender}}, \bibinfo
  {author} {\bibfnamefont {A.}~\bibnamefont {Qaiumzadeh}}, \bibinfo {author}
  {\bibfnamefont {L.}~\bibnamefont {Baldrati}}, \bibinfo {author}
  {\bibfnamefont {J.}~\bibnamefont {Cramer}}, \bibinfo {author} {\bibfnamefont
  {A.}~\bibnamefont {Brataas}}, \bibinfo {author} {\bibfnamefont
  {R.}~\bibnamefont {Duine}},\ and\ \bibinfo {author} {\bibfnamefont
  {M.}~\bibnamefont {Kläui}},\ }\bibfield  {title} {\bibinfo {title} {{Tunable
  long-distance spin transport in a crystalline antiferromagnetic iron
  oxide}},\ }\href@noop {} {\bibfield  {journal} {\bibinfo  {journal} {Nature}\
  }\textbf {\bibinfo {volume} {561}},\ \bibinfo {pages} {222} (\bibinfo {year}
  {2018})}\BibitemShut {NoStop}%
\bibitem [{\citenamefont {Das}\ \emph {et~al.}(2022)\citenamefont {Das},
  \citenamefont {Ross}, \citenamefont {Ma}, \citenamefont {Becker},
  \citenamefont {Schmitt}, \citenamefont {van Duijn}, \citenamefont
  {Galindez-Ruales}, \citenamefont {Fuhrmann}, \citenamefont {Syskaki},
  \citenamefont {Ebels}, \citenamefont {Baltz}, \citenamefont {Barra},
  \citenamefont {Chen}, \citenamefont {Jakob}, \citenamefont {Cao},
  \citenamefont {Sinova}, \citenamefont {Gomonay}, \citenamefont {Lebrun},\
  and\ \citenamefont {Kläui}}]{das}%
  \BibitemOpen
  \bibfield  {author} {\bibinfo {author} {\bibfnamefont {S.}~\bibnamefont
  {Das}}, \bibinfo {author} {\bibfnamefont {A.}~\bibnamefont {Ross}}, \bibinfo
  {author} {\bibfnamefont {X.}~\bibnamefont {Ma}}, \bibinfo {author}
  {\bibfnamefont {S.}~\bibnamefont {Becker}}, \bibinfo {author} {\bibfnamefont
  {C.}~\bibnamefont {Schmitt}}, \bibinfo {author} {\bibfnamefont
  {F.}~\bibnamefont {van Duijn}}, \bibinfo {author} {\bibfnamefont
  {E.}~\bibnamefont {Galindez-Ruales}}, \bibinfo {author} {\bibfnamefont
  {F.}~\bibnamefont {Fuhrmann}}, \bibinfo {author} {\bibfnamefont {M.-A.}\
  \bibnamefont {Syskaki}}, \bibinfo {author} {\bibfnamefont {U.}~\bibnamefont
  {Ebels}}, \bibinfo {author} {\bibfnamefont {V.}~\bibnamefont {Baltz}},
  \bibinfo {author} {\bibfnamefont {A.-L.}\ \bibnamefont {Barra}}, \bibinfo
  {author} {\bibfnamefont {H.}~\bibnamefont {Chen}}, \bibinfo {author}
  {\bibfnamefont {G.}~\bibnamefont {Jakob}}, \bibinfo {author} {\bibfnamefont
  {S.}~\bibnamefont {Cao}}, \bibinfo {author} {\bibfnamefont {J.}~\bibnamefont
  {Sinova}}, \bibinfo {author} {\bibfnamefont {O.}~\bibnamefont {Gomonay}},
  \bibinfo {author} {\bibfnamefont {R.}~\bibnamefont {Lebrun}},\ and\ \bibinfo
  {author} {\bibfnamefont {M.}~\bibnamefont {Kläui}},\ }\bibfield  {title}
  {\bibinfo {title} {{Anisotropic long-range spin transport in canted
  antiferromagnetic orthoferrite $\mathrm{YFeO_3}$}},\ }\href@noop {}
  {\bibfield  {journal} {\bibinfo  {journal} {Nature Communications}\ }\textbf
  {\bibinfo {volume} {13}},\ \bibinfo {pages} {6140} (\bibinfo {year}
  {2022})}\BibitemShut {NoStop}%
\bibitem [{\citenamefont {Hoogeboom}\ \emph {et~al.}(2017)\citenamefont
  {Hoogeboom}, \citenamefont {Aqeel}, \citenamefont {Kuschel}, \citenamefont
  {Palstra},\ and\ \citenamefont {van Wees}}]{hoogeboom2017negative}%
  \BibitemOpen
  \bibfield  {author} {\bibinfo {author} {\bibfnamefont {G.~R.}\ \bibnamefont
  {Hoogeboom}}, \bibinfo {author} {\bibfnamefont {A.}~\bibnamefont {Aqeel}},
  \bibinfo {author} {\bibfnamefont {T.}~\bibnamefont {Kuschel}}, \bibinfo
  {author} {\bibfnamefont {T.~T.}\ \bibnamefont {Palstra}},\ and\ \bibinfo
  {author} {\bibfnamefont {B.~J.}\ \bibnamefont {van Wees}},\ }\bibfield
  {title} {\bibinfo {title} {{Negative spin Hall magnetoresistance of Pt on the
  bulk easy-plane antiferromagnet NiO}},\ }\href@noop {} {\bibfield  {journal}
  {\bibinfo  {journal} {Applied Physics Letters}\ }\textbf {\bibinfo {volume}
  {111}},\ \bibinfo {pages} {052409} (\bibinfo {year} {2017})}\BibitemShut
  {NoStop}%
\bibitem [{\citenamefont {Baldrati}\ \emph {et~al.}(2018)\citenamefont
  {Baldrati}, \citenamefont {Ross}, \citenamefont {Niizeki}, \citenamefont
  {Schneider}, \citenamefont {Ramos}, \citenamefont {Cramer}, \citenamefont
  {Gomonay}, \citenamefont {Filianina}, \citenamefont {Savchenko},
  \citenamefont {Heinze} \emph {et~al.}}]{baldrati2018full}%
  \BibitemOpen
  \bibfield  {author} {\bibinfo {author} {\bibfnamefont {L.}~\bibnamefont
  {Baldrati}}, \bibinfo {author} {\bibfnamefont {A.}~\bibnamefont {Ross}},
  \bibinfo {author} {\bibfnamefont {T.}~\bibnamefont {Niizeki}}, \bibinfo
  {author} {\bibfnamefont {C.}~\bibnamefont {Schneider}}, \bibinfo {author}
  {\bibfnamefont {R.}~\bibnamefont {Ramos}}, \bibinfo {author} {\bibfnamefont
  {J.}~\bibnamefont {Cramer}}, \bibinfo {author} {\bibfnamefont
  {O.}~\bibnamefont {Gomonay}}, \bibinfo {author} {\bibfnamefont
  {M.}~\bibnamefont {Filianina}}, \bibinfo {author} {\bibfnamefont
  {T.}~\bibnamefont {Savchenko}}, \bibinfo {author} {\bibfnamefont
  {D.}~\bibnamefont {Heinze}}, \emph {et~al.},\ }\bibfield  {title} {\bibinfo
  {title} {{Full angular dependence of the spin Hall and ordinary
  magnetoresistance in epitaxial antiferromagnetic NiO (001)/Pt thin films}},\
  }\href@noop {} {\bibfield  {journal} {\bibinfo  {journal} {Physical Review
  B}\ }\textbf {\bibinfo {volume} {98}},\ \bibinfo {pages} {024422} (\bibinfo
  {year} {2018})}\BibitemShut {NoStop}%
\bibitem [{\citenamefont {Wadley}\ \emph {et~al.}(2016)\citenamefont {Wadley},
  \citenamefont {Howells}, \citenamefont {{\v{Z}}elezn{\`y}}, \citenamefont
  {Andrews}, \citenamefont {Hills}, \citenamefont {Campion}, \citenamefont
  {Nov{\'a}k}, \citenamefont {Olejn{\'\i}k}, \citenamefont {Maccherozzi},
  \citenamefont {Dhesi} \emph {et~al.}}]{wadley2016electrical}%
  \BibitemOpen
  \bibfield  {author} {\bibinfo {author} {\bibfnamefont {P.}~\bibnamefont
  {Wadley}}, \bibinfo {author} {\bibfnamefont {B.}~\bibnamefont {Howells}},
  \bibinfo {author} {\bibfnamefont {J.}~\bibnamefont {{\v{Z}}elezn{\`y}}},
  \bibinfo {author} {\bibfnamefont {C.}~\bibnamefont {Andrews}}, \bibinfo
  {author} {\bibfnamefont {V.}~\bibnamefont {Hills}}, \bibinfo {author}
  {\bibfnamefont {R.~P.}\ \bibnamefont {Campion}}, \bibinfo {author}
  {\bibfnamefont {V.}~\bibnamefont {Nov{\'a}k}}, \bibinfo {author}
  {\bibfnamefont {K.}~\bibnamefont {Olejn{\'\i}k}}, \bibinfo {author}
  {\bibfnamefont {F.}~\bibnamefont {Maccherozzi}}, \bibinfo {author}
  {\bibfnamefont {S.}~\bibnamefont {Dhesi}}, \emph {et~al.},\ }\bibfield
  {title} {\bibinfo {title} {{Electrical switching of an antiferromagnet}},\
  }\href@noop {} {\bibfield  {journal} {\bibinfo  {journal} {Science}\ }\textbf
  {\bibinfo {volume} {351}},\ \bibinfo {pages} {587} (\bibinfo {year}
  {2016})}\BibitemShut {NoStop}%
\bibitem [{\citenamefont {Bodnar}\ \emph {et~al.}(2018)\citenamefont {Bodnar},
  \citenamefont {{\v{S}}mejkal}, \citenamefont {Turek}, \citenamefont
  {Jungwirth}, \citenamefont {Gomonay}, \citenamefont {Sinova}, \citenamefont
  {Sapozhnik}, \citenamefont {Elmers}, \citenamefont {Kläui},\ and\
  \citenamefont {Jourdan}}]{bodnar2018writing}%
  \BibitemOpen
  \bibfield  {author} {\bibinfo {author} {\bibfnamefont {S.~Y.}\ \bibnamefont
  {Bodnar}}, \bibinfo {author} {\bibfnamefont {L.}~\bibnamefont
  {{\v{S}}mejkal}}, \bibinfo {author} {\bibfnamefont {I.}~\bibnamefont
  {Turek}}, \bibinfo {author} {\bibfnamefont {T.}~\bibnamefont {Jungwirth}},
  \bibinfo {author} {\bibfnamefont {O.}~\bibnamefont {Gomonay}}, \bibinfo
  {author} {\bibfnamefont {J.}~\bibnamefont {Sinova}}, \bibinfo {author}
  {\bibfnamefont {A.}~\bibnamefont {Sapozhnik}}, \bibinfo {author}
  {\bibfnamefont {H.-J.}\ \bibnamefont {Elmers}}, \bibinfo {author}
  {\bibfnamefont {M.}~\bibnamefont {Kläui}},\ and\ \bibinfo {author}
  {\bibfnamefont {M.}~\bibnamefont {Jourdan}},\ }\bibfield  {title} {\bibinfo
  {title} {{Writing and reading antiferromagnetic $\mathrm{Mn_2Au}$ by N{\'e}el
  spin-orbit torques and large anisotropic magnetoresistance}},\ }\href@noop {}
  {\bibfield  {journal} {\bibinfo  {journal} {Nature Communications}\ }\textbf
  {\bibinfo {volume} {9}},\ \bibinfo {pages} {348} (\bibinfo {year}
  {2018})}\BibitemShut {NoStop}%
\bibitem [{\citenamefont {Lytvynenko}\ \emph {et~al.}(2022)\citenamefont
  {Lytvynenko}, \citenamefont {Reimers}, \citenamefont {Niu}, \citenamefont
  {Golias}, \citenamefont {Sarpi}, \citenamefont {Ishibe-Veiga}, \citenamefont
  {Denneulin}, \citenamefont {Kovacs}, \citenamefont {Dunin-Borkowski},
  \citenamefont {Kläui} \emph {et~al.}}]{lytvynenko2022current}%
  \BibitemOpen
  \bibfield  {author} {\bibinfo {author} {\bibfnamefont {Y.}~\bibnamefont
  {Lytvynenko}}, \bibinfo {author} {\bibfnamefont {S.}~\bibnamefont {Reimers}},
  \bibinfo {author} {\bibfnamefont {Y.}~\bibnamefont {Niu}}, \bibinfo {author}
  {\bibfnamefont {E.}~\bibnamefont {Golias}}, \bibinfo {author} {\bibfnamefont
  {B.}~\bibnamefont {Sarpi}}, \bibinfo {author} {\bibfnamefont
  {L.}~\bibnamefont {Ishibe-Veiga}}, \bibinfo {author} {\bibfnamefont
  {T.}~\bibnamefont {Denneulin}}, \bibinfo {author} {\bibfnamefont
  {A.}~\bibnamefont {Kovacs}}, \bibinfo {author} {\bibfnamefont
  {R.}~\bibnamefont {Dunin-Borkowski}}, \bibinfo {author} {\bibfnamefont
  {M.}~\bibnamefont {Kläui}}, \emph {et~al.},\ }\bibfield  {title} {\bibinfo
  {title} {{Current-driven writing process in antiferromagnetic
  $\mathrm{Mn_2Au}$ for memory applications}},\ }\href@noop {} {\bibfield
  {journal} {\bibinfo  {journal} {arXiv preprint arXiv:2208.04048}\ } (\bibinfo
  {year} {2022})}\BibitemShut {NoStop}%
\bibitem [{\citenamefont {Moriyama}\ \emph {et~al.}(2018)\citenamefont
  {Moriyama}, \citenamefont {Oda}, \citenamefont {Ohkochi}, \citenamefont
  {Kimata},\ and\ \citenamefont {Ono}}]{moriyama2018spin}%
  \BibitemOpen
  \bibfield  {author} {\bibinfo {author} {\bibfnamefont {T.}~\bibnamefont
  {Moriyama}}, \bibinfo {author} {\bibfnamefont {K.}~\bibnamefont {Oda}},
  \bibinfo {author} {\bibfnamefont {T.}~\bibnamefont {Ohkochi}}, \bibinfo
  {author} {\bibfnamefont {M.}~\bibnamefont {Kimata}},\ and\ \bibinfo {author}
  {\bibfnamefont {T.}~\bibnamefont {Ono}},\ }\bibfield  {title} {\bibinfo
  {title} {{Spin torque control of antiferromagnetic moments in NiO}},\
  }\href@noop {} {\bibfield  {journal} {\bibinfo  {journal} {Scientific
  Reports}\ }\textbf {\bibinfo {volume} {8}},\ \bibinfo {pages} {14167}
  (\bibinfo {year} {2018})}\BibitemShut {NoStop}%
\bibitem [{\citenamefont {Chen}\ \emph {et~al.}(2018)\citenamefont {Chen},
  \citenamefont {Zarzuela}, \citenamefont {Zhang}, \citenamefont {Song},
  \citenamefont {Zhou}, \citenamefont {Shi}, \citenamefont {Li}, \citenamefont
  {Zhou}, \citenamefont {Jiang}, \citenamefont {Pan} \emph
  {et~al.}}]{chen2018antidamping}%
  \BibitemOpen
  \bibfield  {author} {\bibinfo {author} {\bibfnamefont {X.}~\bibnamefont
  {Chen}}, \bibinfo {author} {\bibfnamefont {R.}~\bibnamefont {Zarzuela}},
  \bibinfo {author} {\bibfnamefont {J.}~\bibnamefont {Zhang}}, \bibinfo
  {author} {\bibfnamefont {C.}~\bibnamefont {Song}}, \bibinfo {author}
  {\bibfnamefont {X.}~\bibnamefont {Zhou}}, \bibinfo {author} {\bibfnamefont
  {G.}~\bibnamefont {Shi}}, \bibinfo {author} {\bibfnamefont {F.}~\bibnamefont
  {Li}}, \bibinfo {author} {\bibfnamefont {H.}~\bibnamefont {Zhou}}, \bibinfo
  {author} {\bibfnamefont {W.}~\bibnamefont {Jiang}}, \bibinfo {author}
  {\bibfnamefont {F.}~\bibnamefont {Pan}}, \emph {et~al.},\ }\bibfield  {title}
  {\bibinfo {title} {{Antidamping-torque-induced switching in biaxial
  antiferromagnetic insulators}},\ }\href@noop {} {\bibfield  {journal}
  {\bibinfo  {journal} {Physical Review Letters}\ }\textbf {\bibinfo {volume}
  {120}},\ \bibinfo {pages} {207204} (\bibinfo {year} {2018})}\BibitemShut
  {NoStop}%
\bibitem [{\citenamefont {Baldrati}\ \emph {et~al.}(2019)\citenamefont
  {Baldrati}, \citenamefont {Gomonay}, \citenamefont {Ross}, \citenamefont
  {Filianina}, \citenamefont {Lebrun}, \citenamefont {Ramos}, \citenamefont
  {Leveille}, \citenamefont {Fuhrmann}, \citenamefont {Forrest}, \citenamefont
  {Maccherozzi} \emph {et~al.}}]{baldrati2019mechanism}%
  \BibitemOpen
  \bibfield  {author} {\bibinfo {author} {\bibfnamefont {L.}~\bibnamefont
  {Baldrati}}, \bibinfo {author} {\bibfnamefont {O.}~\bibnamefont {Gomonay}},
  \bibinfo {author} {\bibfnamefont {A.}~\bibnamefont {Ross}}, \bibinfo {author}
  {\bibfnamefont {M.}~\bibnamefont {Filianina}}, \bibinfo {author}
  {\bibfnamefont {R.}~\bibnamefont {Lebrun}}, \bibinfo {author} {\bibfnamefont
  {R.}~\bibnamefont {Ramos}}, \bibinfo {author} {\bibfnamefont
  {C.}~\bibnamefont {Leveille}}, \bibinfo {author} {\bibfnamefont
  {F.}~\bibnamefont {Fuhrmann}}, \bibinfo {author} {\bibfnamefont
  {T.}~\bibnamefont {Forrest}}, \bibinfo {author} {\bibfnamefont
  {F.}~\bibnamefont {Maccherozzi}}, \emph {et~al.},\ }\bibfield  {title}
  {\bibinfo {title} {{Mechanism of N{\'e}el order switching in
  antiferromagnetic thin films revealed by magnetotransport and direct
  imaging}},\ }\href@noop {} {\bibfield  {journal} {\bibinfo  {journal}
  {Physical Review Letters}\ }\textbf {\bibinfo {volume} {123}},\ \bibinfo
  {pages} {177201} (\bibinfo {year} {2019})}\BibitemShut {NoStop}%
\bibitem [{\citenamefont {Zhang}\ \emph {et~al.}(2019)\citenamefont {Zhang},
  \citenamefont {Finley}, \citenamefont {Safi},\ and\ \citenamefont
  {Liu}}]{zhang2019quantitative}%
  \BibitemOpen
  \bibfield  {author} {\bibinfo {author} {\bibfnamefont {P.}~\bibnamefont
  {Zhang}}, \bibinfo {author} {\bibfnamefont {J.}~\bibnamefont {Finley}},
  \bibinfo {author} {\bibfnamefont {T.}~\bibnamefont {Safi}},\ and\ \bibinfo
  {author} {\bibfnamefont {L.}~\bibnamefont {Liu}},\ }\bibfield  {title}
  {\bibinfo {title} {{Quantitative study on current-induced effect in an
  antiferromagnet insulator/Pt bilayer film}},\ }\href@noop {} {\bibfield
  {journal} {\bibinfo  {journal} {Physical Review Letters}\ }\textbf {\bibinfo
  {volume} {123}},\ \bibinfo {pages} {247206} (\bibinfo {year}
  {2019})}\BibitemShut {NoStop}%
\bibitem [{\citenamefont {Meer}\ \emph {et~al.}(2020)\citenamefont {Meer},
  \citenamefont {Schreiber}, \citenamefont {Schmitt}, \citenamefont {Ramos},
  \citenamefont {Saitoh}, \citenamefont {Gomonay}, \citenamefont {Sinova},
  \citenamefont {Baldrati},\ and\ \citenamefont {Kläui}}]{meer2020direct}%
  \BibitemOpen
  \bibfield  {author} {\bibinfo {author} {\bibfnamefont {H.}~\bibnamefont
  {Meer}}, \bibinfo {author} {\bibfnamefont {F.}~\bibnamefont {Schreiber}},
  \bibinfo {author} {\bibfnamefont {C.}~\bibnamefont {Schmitt}}, \bibinfo
  {author} {\bibfnamefont {R.}~\bibnamefont {Ramos}}, \bibinfo {author}
  {\bibfnamefont {E.}~\bibnamefont {Saitoh}}, \bibinfo {author} {\bibfnamefont
  {O.}~\bibnamefont {Gomonay}}, \bibinfo {author} {\bibfnamefont
  {J.}~\bibnamefont {Sinova}}, \bibinfo {author} {\bibfnamefont
  {L.}~\bibnamefont {Baldrati}},\ and\ \bibinfo {author} {\bibfnamefont
  {M.}~\bibnamefont {Kläui}},\ }\bibfield  {title} {\bibinfo {title} {{Direct
  imaging of current-induced antiferromagnetic switching revealing a pure
  thermomagnetoelastic switching mechanism in NiO}},\ }\href@noop {} {\bibfield
   {journal} {\bibinfo  {journal} {Nano Letters}\ }\textbf {\bibinfo {volume}
  {21}},\ \bibinfo {pages} {114} (\bibinfo {year} {2020})}\BibitemShut
  {NoStop}%
\bibitem [{\citenamefont {Baldrati}\ \emph {et~al.}(2020)\citenamefont
  {Baldrati}, \citenamefont {Schmitt}, \citenamefont {Gomonay}, \citenamefont
  {Lebrun}, \citenamefont {Ramos}, \citenamefont {Saitoh}, \citenamefont
  {Sinova},\ and\ \citenamefont {Kläui}}]{baldrati2020efficient}%
  \BibitemOpen
  \bibfield  {author} {\bibinfo {author} {\bibfnamefont {L.}~\bibnamefont
  {Baldrati}}, \bibinfo {author} {\bibfnamefont {C.}~\bibnamefont {Schmitt}},
  \bibinfo {author} {\bibfnamefont {O.}~\bibnamefont {Gomonay}}, \bibinfo
  {author} {\bibfnamefont {R.}~\bibnamefont {Lebrun}}, \bibinfo {author}
  {\bibfnamefont {R.}~\bibnamefont {Ramos}}, \bibinfo {author} {\bibfnamefont
  {E.}~\bibnamefont {Saitoh}}, \bibinfo {author} {\bibfnamefont
  {J.}~\bibnamefont {Sinova}},\ and\ \bibinfo {author} {\bibfnamefont
  {M.}~\bibnamefont {Kläui}},\ }\bibfield  {title} {\bibinfo {title}
  {{Efficient spin torques in antiferromagnetic CoO/Pt quantified by comparing
  field-and current-induced switching}},\ }\href@noop {} {\bibfield  {journal}
  {\bibinfo  {journal} {Physical Review Letters}\ }\textbf {\bibinfo {volume}
  {125}},\ \bibinfo {pages} {077201} (\bibinfo {year} {2020})}\BibitemShut
  {NoStop}%
\bibitem [{\citenamefont {Gray}\ \emph {et~al.}(2019)\citenamefont {Gray},
  \citenamefont {Moriyama}, \citenamefont {Sivadas}, \citenamefont {Stiehl},
  \citenamefont {Heron}, \citenamefont {Need}, \citenamefont {Kirby},
  \citenamefont {Low}, \citenamefont {Nowack}, \citenamefont {Schlom} \emph
  {et~al.}}]{gray2019spin}%
  \BibitemOpen
  \bibfield  {author} {\bibinfo {author} {\bibfnamefont {I.}~\bibnamefont
  {Gray}}, \bibinfo {author} {\bibfnamefont {T.}~\bibnamefont {Moriyama}},
  \bibinfo {author} {\bibfnamefont {N.}~\bibnamefont {Sivadas}}, \bibinfo
  {author} {\bibfnamefont {G.~M.}\ \bibnamefont {Stiehl}}, \bibinfo {author}
  {\bibfnamefont {J.~T.}\ \bibnamefont {Heron}}, \bibinfo {author}
  {\bibfnamefont {R.}~\bibnamefont {Need}}, \bibinfo {author} {\bibfnamefont
  {B.~J.}\ \bibnamefont {Kirby}}, \bibinfo {author} {\bibfnamefont {D.~H.}\
  \bibnamefont {Low}}, \bibinfo {author} {\bibfnamefont {K.~C.}\ \bibnamefont
  {Nowack}}, \bibinfo {author} {\bibfnamefont {D.~G.}\ \bibnamefont {Schlom}},
  \emph {et~al.},\ }\bibfield  {title} {\bibinfo {title} {{Spin Seebeck imaging
  of spin-torque switching in antiferromagnetic Pt/NiO heterostructures}},\
  }\href@noop {} {\bibfield  {journal} {\bibinfo  {journal} {Physical Review
  X}\ }\textbf {\bibinfo {volume} {9}},\ \bibinfo {pages} {041016} (\bibinfo
  {year} {2019})}\BibitemShut {NoStop}%
\bibitem [{\citenamefont {Cheng}\ \emph {et~al.}(2020)\citenamefont {Cheng},
  \citenamefont {Yu}, \citenamefont {Zhu}, \citenamefont {Hwang},\ and\
  \citenamefont {Yang}}]{cheng2020electrical}%
  \BibitemOpen
  \bibfield  {author} {\bibinfo {author} {\bibfnamefont {Y.}~\bibnamefont
  {Cheng}}, \bibinfo {author} {\bibfnamefont {S.}~\bibnamefont {Yu}}, \bibinfo
  {author} {\bibfnamefont {M.}~\bibnamefont {Zhu}}, \bibinfo {author}
  {\bibfnamefont {J.}~\bibnamefont {Hwang}},\ and\ \bibinfo {author}
  {\bibfnamefont {F.}~\bibnamefont {Yang}},\ }\bibfield  {title} {\bibinfo
  {title} {{Electrical Switching of Tristate Antiferromagnetic N{\'e}el Order
  in $\alpha - \mathrm{Fe_2O_3}$ Epitaxial Films}},\ }\href@noop {} {\bibfield
  {journal} {\bibinfo  {journal} {Physical Review Letters}\ }\textbf {\bibinfo
  {volume} {124}},\ \bibinfo {pages} {027202} (\bibinfo {year}
  {2020})}\BibitemShut {NoStop}%
\bibitem [{\citenamefont {Zhang}\ \emph {et~al.}(2022)\citenamefont {Zhang},
  \citenamefont {Chou}, \citenamefont {Yun}, \citenamefont {McGoldrick},
  \citenamefont {Hou}, \citenamefont {Mkhoyan},\ and\ \citenamefont
  {Liu}}]{zhang2022control}%
  \BibitemOpen
  \bibfield  {author} {\bibinfo {author} {\bibfnamefont {P.}~\bibnamefont
  {Zhang}}, \bibinfo {author} {\bibfnamefont {C.-T.}\ \bibnamefont {Chou}},
  \bibinfo {author} {\bibfnamefont {H.}~\bibnamefont {Yun}}, \bibinfo {author}
  {\bibfnamefont {B.~C.}\ \bibnamefont {McGoldrick}}, \bibinfo {author}
  {\bibfnamefont {J.~T.}\ \bibnamefont {Hou}}, \bibinfo {author} {\bibfnamefont
  {K.~A.}\ \bibnamefont {Mkhoyan}},\ and\ \bibinfo {author} {\bibfnamefont
  {L.}~\bibnamefont {Liu}},\ }\bibfield  {title} {\bibinfo {title} {{Control of
  N{\'e}el Vector with Spin-Orbit Torques in an Antiferromagnetic Insulator
  with Tilted Easy Plane}},\ }\href@noop {} {\bibfield  {journal} {\bibinfo
  {journal} {Physical Review Letters}\ }\textbf {\bibinfo {volume} {129}},\
  \bibinfo {pages} {017203} (\bibinfo {year} {2022})}\BibitemShut {NoStop}%
\bibitem [{\citenamefont {Roth}(1958)}]{roth1958magnetic}%
  \BibitemOpen
  \bibfield  {author} {\bibinfo {author} {\bibfnamefont {W.}~\bibnamefont
  {Roth}},\ }\bibfield  {title} {\bibinfo {title} {{Magnetic structures of MnO,
  FeO, CoO, and NiO}},\ }\href@noop {} {\bibfield  {journal} {\bibinfo
  {journal} {Physical Review}\ }\textbf {\bibinfo {volume} {110}},\ \bibinfo
  {pages} {1333} (\bibinfo {year} {1958})}\BibitemShut {NoStop}%
\bibitem [{\citenamefont {Uchida}\ \emph {et~al.}(1964)\citenamefont {Uchida},
  \citenamefont {Fukuoka}, \citenamefont {Kondoh}, \citenamefont {Takeda},
  \citenamefont {Nakazumi},\ and\ \citenamefont
  {Nagamiya}}]{uchida1964magnetic}%
  \BibitemOpen
  \bibfield  {author} {\bibinfo {author} {\bibfnamefont {E.}~\bibnamefont
  {Uchida}}, \bibinfo {author} {\bibfnamefont {N.}~\bibnamefont {Fukuoka}},
  \bibinfo {author} {\bibfnamefont {H.}~\bibnamefont {Kondoh}}, \bibinfo
  {author} {\bibfnamefont {T.}~\bibnamefont {Takeda}}, \bibinfo {author}
  {\bibfnamefont {Y.}~\bibnamefont {Nakazumi}},\ and\ \bibinfo {author}
  {\bibfnamefont {T.}~\bibnamefont {Nagamiya}},\ }\bibfield  {title} {\bibinfo
  {title} {{Magnetic anisotropy measurements of CoO single crystal}},\
  }\href@noop {} {\bibfield  {journal} {\bibinfo  {journal} {Journal of the
  Physical Society of Japan}\ }\textbf {\bibinfo {volume} {19}},\ \bibinfo
  {pages} {2088} (\bibinfo {year} {1964})}\BibitemShut {NoStop}%
\bibitem [{\citenamefont {Saito}\ \emph {et~al.}(1966)\citenamefont {Saito},
  \citenamefont {Nakahigashi},\ and\ \citenamefont {Shimomura}}]{saito1966x}%
  \BibitemOpen
  \bibfield  {author} {\bibinfo {author} {\bibfnamefont {S.}~\bibnamefont
  {Saito}}, \bibinfo {author} {\bibfnamefont {K.}~\bibnamefont {Nakahigashi}},\
  and\ \bibinfo {author} {\bibfnamefont {Y.}~\bibnamefont {Shimomura}},\
  }\bibfield  {title} {\bibinfo {title} {{X-ray diffraction study on CoO}},\
  }\href@noop {} {\bibfield  {journal} {\bibinfo  {journal} {Journal of the
  Physical Society of Japan}\ }\textbf {\bibinfo {volume} {21}},\ \bibinfo
  {pages} {850} (\bibinfo {year} {1966})}\BibitemShut {NoStop}%
\bibitem [{\citenamefont {Csiszar}\ \emph {et~al.}(2005)\citenamefont
  {Csiszar}, \citenamefont {Haverkort}, \citenamefont {Hu}, \citenamefont
  {Tanaka}, \citenamefont {Hsieh}, \citenamefont {Lin}, \citenamefont {Chen},
  \citenamefont {Hibma},\ and\ \citenamefont {Tjeng}}]{csiszar2005controlling}%
  \BibitemOpen
  \bibfield  {author} {\bibinfo {author} {\bibfnamefont {S.}~\bibnamefont
  {Csiszar}}, \bibinfo {author} {\bibfnamefont {M.}~\bibnamefont {Haverkort}},
  \bibinfo {author} {\bibfnamefont {Z.}~\bibnamefont {Hu}}, \bibinfo {author}
  {\bibfnamefont {A.}~\bibnamefont {Tanaka}}, \bibinfo {author} {\bibfnamefont
  {H.}~\bibnamefont {Hsieh}}, \bibinfo {author} {\bibfnamefont {H.-J.}\
  \bibnamefont {Lin}}, \bibinfo {author} {\bibfnamefont {C.}~\bibnamefont
  {Chen}}, \bibinfo {author} {\bibfnamefont {T.}~\bibnamefont {Hibma}},\ and\
  \bibinfo {author} {\bibfnamefont {L.}~\bibnamefont {Tjeng}},\ }\bibfield
  {title} {\bibinfo {title} {{Controlling orbital moment and spin orientation
  in CoO layers by strain}},\ }\href@noop {} {\bibfield  {journal} {\bibinfo
  {journal} {Physical Review Letters}\ }\textbf {\bibinfo {volume} {95}},\
  \bibinfo {pages} {187205} (\bibinfo {year} {2005})}\BibitemShut {NoStop}%
\bibitem [{\citenamefont {Zhu}\ \emph {et~al.}(2014)\citenamefont {Zhu},
  \citenamefont {Li}, \citenamefont {Li}, \citenamefont {Ding}, \citenamefont
  {Liang}, \citenamefont {Xiao}, \citenamefont {Luo}, \citenamefont {Hua},
  \citenamefont {Lin}, \citenamefont {Pi} \emph
  {et~al.}}]{zhu2014antiferromagnetic}%
  \BibitemOpen
  \bibfield  {author} {\bibinfo {author} {\bibfnamefont {J.}~\bibnamefont
  {Zhu}}, \bibinfo {author} {\bibfnamefont {Q.}~\bibnamefont {Li}}, \bibinfo
  {author} {\bibfnamefont {J.}~\bibnamefont {Li}}, \bibinfo {author}
  {\bibfnamefont {Z.}~\bibnamefont {Ding}}, \bibinfo {author} {\bibfnamefont
  {J.}~\bibnamefont {Liang}}, \bibinfo {author} {\bibfnamefont
  {X.}~\bibnamefont {Xiao}}, \bibinfo {author} {\bibfnamefont {Y.}~\bibnamefont
  {Luo}}, \bibinfo {author} {\bibfnamefont {C.}~\bibnamefont {Hua}}, \bibinfo
  {author} {\bibfnamefont {H.-J.}\ \bibnamefont {Lin}}, \bibinfo {author}
  {\bibfnamefont {T.}~\bibnamefont {Pi}}, \emph {et~al.},\ }\bibfield  {title}
  {\bibinfo {title} {{Antiferromagnetic spin reorientation transition in
  epitaxial NiO/CoO/MgO (001) systems}},\ }\href@noop {} {\bibfield  {journal}
  {\bibinfo  {journal} {Physical Review B}\ }\textbf {\bibinfo {volume} {90}},\
  \bibinfo {pages} {054403} (\bibinfo {year} {2014})}\BibitemShut {NoStop}%
\bibitem [{\citenamefont {Cao}\ \emph {et~al.}(2011)\citenamefont {Cao},
  \citenamefont {Li}, \citenamefont {Chen}, \citenamefont {Zhu}, \citenamefont
  {Hu},\ and\ \citenamefont {Wu}}]{cao2011temperature}%
  \BibitemOpen
  \bibfield  {author} {\bibinfo {author} {\bibfnamefont {W.}~\bibnamefont
  {Cao}}, \bibinfo {author} {\bibfnamefont {J.}~\bibnamefont {Li}}, \bibinfo
  {author} {\bibfnamefont {G.}~\bibnamefont {Chen}}, \bibinfo {author}
  {\bibfnamefont {J.}~\bibnamefont {Zhu}}, \bibinfo {author} {\bibfnamefont
  {C.}~\bibnamefont {Hu}},\ and\ \bibinfo {author} {\bibfnamefont
  {Y.}~\bibnamefont {Wu}},\ }\bibfield  {title} {\bibinfo {title}
  {{Temperature-dependent magnetic anisotropies in epitaxial Fe/CoO/MgO (001)
  system studied by the planar Hall effect}},\ }\href@noop {} {\bibfield
  {journal} {\bibinfo  {journal} {Applied Physics Letters}\ }\textbf {\bibinfo
  {volume} {98}},\ \bibinfo {pages} {262506} (\bibinfo {year}
  {2011})}\BibitemShut {NoStop}%
\bibitem [{\citenamefont {Grzybowski}\ \emph {et~al.}(2023)\citenamefont
  {Grzybowski}, \citenamefont {Schippers}, \citenamefont {Gomonay},
  \citenamefont {Rubi}, \citenamefont {Bal}, \citenamefont {Zeitler},
  \citenamefont {Kozio{\l}-Rachwa{\l}}, \citenamefont {Szpytma}, \citenamefont
  {Janus}, \citenamefont {Kurowska} \emph
  {et~al.}}]{grzybowski2023antiferromagnetic}%
  \BibitemOpen
  \bibfield  {author} {\bibinfo {author} {\bibfnamefont {M.}~\bibnamefont
  {Grzybowski}}, \bibinfo {author} {\bibfnamefont {C.}~\bibnamefont
  {Schippers}}, \bibinfo {author} {\bibfnamefont {O.}~\bibnamefont {Gomonay}},
  \bibinfo {author} {\bibfnamefont {K.}~\bibnamefont {Rubi}}, \bibinfo {author}
  {\bibfnamefont {M.}~\bibnamefont {Bal}}, \bibinfo {author} {\bibfnamefont
  {U.}~\bibnamefont {Zeitler}}, \bibinfo {author} {\bibfnamefont
  {A.}~\bibnamefont {Kozio{\l}-Rachwa{\l}}}, \bibinfo {author} {\bibfnamefont
  {M.}~\bibnamefont {Szpytma}}, \bibinfo {author} {\bibfnamefont
  {W.}~\bibnamefont {Janus}}, \bibinfo {author} {\bibfnamefont
  {B.}~\bibnamefont {Kurowska}}, \emph {et~al.},\ }\bibfield  {title} {\bibinfo
  {title} {{Antiferromagnetic hysteresis above the spin-flop field}},\
  }\href@noop {} {\bibfield  {journal} {\bibinfo  {journal} {Physical Review
  B}\ }\textbf {\bibinfo {volume} {107}},\ \bibinfo {pages} {L060403} (\bibinfo
  {year} {2023})}\BibitemShut {NoStop}%
\bibitem [{\citenamefont {Nakayama}\ \emph {et~al.}(2013)\citenamefont
  {Nakayama}, \citenamefont {Althammer}, \citenamefont {Chen}, \citenamefont
  {Uchida}, \citenamefont {Kajiwara}, \citenamefont {Kikuchi}, \citenamefont
  {Ohtani}, \citenamefont {Geprägs}, \citenamefont {Opel}, \citenamefont
  {Takahashi} \emph {et~al.}}]{nakayama2013spin}%
  \BibitemOpen
  \bibfield  {author} {\bibinfo {author} {\bibfnamefont {H.}~\bibnamefont
  {Nakayama}}, \bibinfo {author} {\bibfnamefont {M.}~\bibnamefont {Althammer}},
  \bibinfo {author} {\bibfnamefont {Y.-T.}\ \bibnamefont {Chen}}, \bibinfo
  {author} {\bibfnamefont {K.-i.}\ \bibnamefont {Uchida}}, \bibinfo {author}
  {\bibfnamefont {Y.}~\bibnamefont {Kajiwara}}, \bibinfo {author}
  {\bibfnamefont {D.}~\bibnamefont {Kikuchi}}, \bibinfo {author} {\bibfnamefont
  {T.}~\bibnamefont {Ohtani}}, \bibinfo {author} {\bibfnamefont
  {S.}~\bibnamefont {Geprägs}}, \bibinfo {author} {\bibfnamefont
  {M.}~\bibnamefont {Opel}}, \bibinfo {author} {\bibfnamefont {S.}~\bibnamefont
  {Takahashi}}, \emph {et~al.},\ }\bibfield  {title} {\bibinfo {title} {{Spin
  Hall magnetoresistance induced by a nonequilibrium proximity effect}},\
  }\href@noop {} {\bibfield  {journal} {\bibinfo  {journal} {Physical Review
  Letters}\ }\textbf {\bibinfo {volume} {110}},\ \bibinfo {pages} {206601}
  (\bibinfo {year} {2013})}\BibitemShut {NoStop}%
\bibitem [{\citenamefont {Schreiber}\ \emph {et~al.}(2021)\citenamefont
  {Schreiber}, \citenamefont {Meer}, \citenamefont {Schmitt}, \citenamefont
  {Ramos}, \citenamefont {Saitoh}, \citenamefont {Baldrati},\ and\
  \citenamefont {Kläui}}]{schreiber2021magnetic}%
  \BibitemOpen
  \bibfield  {author} {\bibinfo {author} {\bibfnamefont {F.}~\bibnamefont
  {Schreiber}}, \bibinfo {author} {\bibfnamefont {H.}~\bibnamefont {Meer}},
  \bibinfo {author} {\bibfnamefont {C.}~\bibnamefont {Schmitt}}, \bibinfo
  {author} {\bibfnamefont {R.}~\bibnamefont {Ramos}}, \bibinfo {author}
  {\bibfnamefont {E.}~\bibnamefont {Saitoh}}, \bibinfo {author} {\bibfnamefont
  {L.}~\bibnamefont {Baldrati}},\ and\ \bibinfo {author} {\bibfnamefont
  {M.}~\bibnamefont {Kläui}},\ }\bibfield  {title} {\bibinfo {title}
  {{Magnetic sensitivity distribution of Hall devices in antiferromagnetic
  switching experiments}},\ }\href@noop {} {\bibfield  {journal} {\bibinfo
  {journal} {Physical Review Applied}\ }\textbf {\bibinfo {volume} {16}},\
  \bibinfo {pages} {064023} (\bibinfo {year} {2021})}\BibitemShut {NoStop}%
\bibitem [{\citenamefont {Schreiber}\ \emph {et~al.}(2020)\citenamefont
  {Schreiber}, \citenamefont {Baldrati}, \citenamefont {Schmitt}, \citenamefont
  {Ramos}, \citenamefont {Saitoh}, \citenamefont {Lebrun},\ and\ \citenamefont
  {Kläui}}]{schreiber2020concurrent}%
  \BibitemOpen
  \bibfield  {author} {\bibinfo {author} {\bibfnamefont {F.}~\bibnamefont
  {Schreiber}}, \bibinfo {author} {\bibfnamefont {L.}~\bibnamefont {Baldrati}},
  \bibinfo {author} {\bibfnamefont {C.}~\bibnamefont {Schmitt}}, \bibinfo
  {author} {\bibfnamefont {R.}~\bibnamefont {Ramos}}, \bibinfo {author}
  {\bibfnamefont {E.}~\bibnamefont {Saitoh}}, \bibinfo {author} {\bibfnamefont
  {R.}~\bibnamefont {Lebrun}},\ and\ \bibinfo {author} {\bibfnamefont
  {M.}~\bibnamefont {Kläui}},\ }\bibfield  {title} {\bibinfo {title}
  {{Concurrent magneto-optical imaging and magneto-transport readout of
  electrical switching of insulating antiferromagnetic thin films}},\
  }\href@noop {} {\bibfield  {journal} {\bibinfo  {journal} {Applied Physics
  Letters}\ }\textbf {\bibinfo {volume} {117}},\ \bibinfo {pages} {082401}
  (\bibinfo {year} {2020})}\BibitemShut {NoStop}%
\end{thebibliography}%

\section{Supporting information}
\subsection{Electrical switching of MgO(001)//CoO(2 nm)/Pt(2 nm) bilayers}
\begin{figure*}
 \includegraphics[width=16cm]{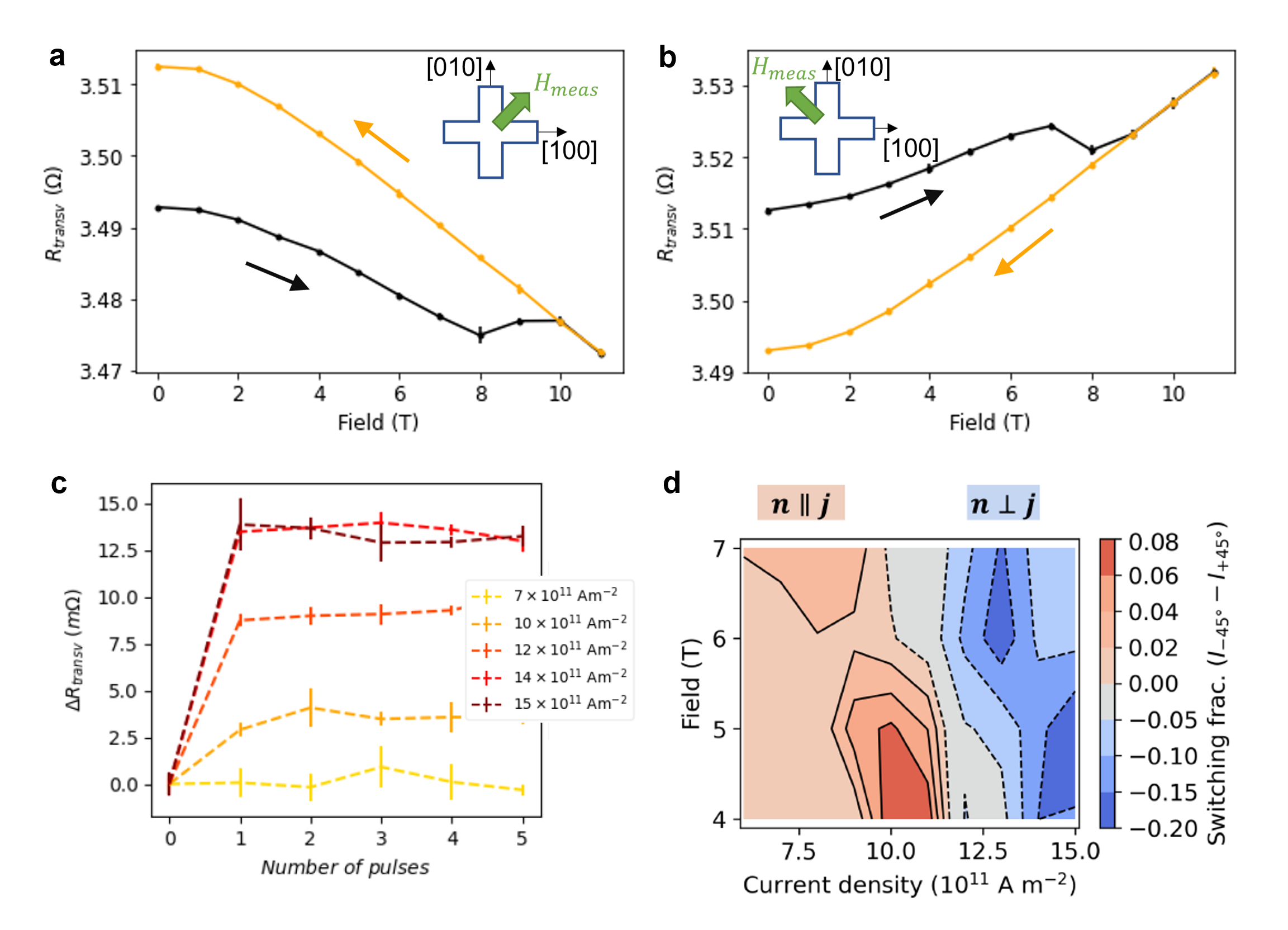}
 \caption{(a) Field-induced spin flop at $\SI{200}{K}$, read by SMR in the presence of a field applied along the [110] direction in CoO($\SI{2}{nm}$)/Pt($\SI{2}{nm}$). A $\SI{11}{T}$ field was previously applied along the orthogonal direction to set a well-defined starting state. (b) Field-induced spin flop with orthogonal field direction ($[\bar{1}10]$) compared to the previous scan. (c) Transverse resistance variation versus pulse current density, probing the threshold and saturation of the switching. Before the measurements, a reset field ($\mu_0 \mathbf{H}_{\mathrm{reset}} = \SI{11}{T}$) was applied along $[\bar{1}10]$, followed by five pulses along +45° direction while a static magnetic field of $\SI{7}{T}$ was applied along [110]. (d) Difference of the switching fraction after -45°- and +45°-pulses as a function of the applied field and pulse current density. The lines are contour plots with constant switching efficiency.}
 \label{fig:three}
\end{figure*}

We investigate switching in a CoO($\SI{2}{nm}$) sample by transport measurements. First, field sweep measurements along the two easy axes were performed. In order to have a well defined starting state with $\mathbf{n} \parallel [110]$, a magnetic field of $\mu_0 \mathbf{H} = \SI{11}{T}$ was applied along $[\bar{1}10]$. The subsequent field sweep along the [110] direction up to a field of $\mu_0 \mathbf{H} = \SI{11}{T}$ at $T = \SI{200}{K}$ is shown in Fig. \ref{fig:three} (a). One can observe a spin-flop transition for a applied magnetic field of $\mu_0 \mathbf{H} = \SI{8}{T}-\SI{10}{T}$. In a second step a field sweep along $[\bar{1}10]$ direction was carried out. As shown in Fig. \ref{fig:three} (b) the spin flop transition along this axis is at fields of $\mu_0 \mathbf{H} = \SI{7}{T}-\SI{9}{T}$. The difference in spin flop fields shows that the alignment of the Néel vector along [110] is easier than an alignment along $[\bar{1}10]$. This indicates a cubic anisotropy defining two easy axes with a superimposed uniaxial anisotropy. Due to this, a reversible switching between two perpendicular states of the Néel vector is hindered. Instead, we use a switching geometry where we combine current-induced switching with the application of a magnetic field of varying strength. With a field applied along [110] during the application of the electric pulses, switching of the Néel vector away from the preferred direction can be facilitated. In Fig. \ref{fig:three} (c) the transverse resistance variation $R_{\mathrm{transv}}$ is shown as a function of the number of pulses for +45°-pulses of various switching current densities while a static magnetic field of $\SI{7}{T}$ was applied along [110]. Before each measurement the system was prepared in the same reproducible starting state by applying a magnetic field of $\SI{11}{T}$ along $[\bar{1}10]$ and all measurements were performed at $T = \SI{205}{K}$. Below the threshold current ($\mathbf{j} = \SI{7e11}{Am}^{-2}$) no change in resistance is observed. Increasing the pulse current density leads to an increasing change in $R_{\mathrm{transv}}$, corresponding to a switching of $\mathbf{n}$ away from the [110] direction. Field sweeps along the two easy axes yield the change in transverse resistance that corresponds to a full switching of the whole sample from one easy axis to another. This can be used to determine the switching fraction in electrical measurements assuming it is proportional to the resistance increase. Taking this into account, one can now perform switching along the -45° and +45° directions and compare the switching fractions. One must note, that -45°- and +45°-pulses are expected to result in orthogonal states of the Néel vector. However, there is the additional contribution of the static magnetic field to the switching. With increasing pulse current density, the system is heated up and the switching effect of the magnetic field becomes stronger, always favoring the Néel vector to be oriented perpendicular to the applied field. Therefore, with increasing current density the switching fraction will increase for both, -45°- as well as +45°-pulses. However, a slight difference in switching fraction will yield information of the switching sign that is favored by the current pulses. The pulsing direction which favors the same Néel vector orientation as the magnetic field will yield slightly higher switching fractions. To compare the effects of the different pulsing directions, in Fig. \ref{fig:three} (d) the difference in switching fractions $(I_{-45°}-I_{+45°})$ is plotted as a function of pulse current density and magnetic field. One can see that this, as already observed for thicker CoO($\SI{4}{nm}$) films, reveals two different switching regimes. At low current densities -45°-pulses are leading to larger switching fractions compared to +45°-pulses (red) and are more effective under otherwise identical conditions. This indicates that the preferred alignment of $\mathbf{n}$ in a low current density regime is parallel to the applied current pulse direction ($\mathbf{n}\parallel \mathbf{j}$). For higher current densities, however, +45°-pulses are more effective compared to -45°-pulses (blue). Meaning that in the high current density regime an orientation of the Néel vector perpendicular to the pulse direction is preferred ($\mathbf{n} \perp \mathbf{j}$).

\end{document}